\documentclass[aps, prl, reprint, superscriptaddress]{revtex4-2}

\usepackage{graphicx}
\usepackage{dcolumn}
\usepackage{bm}
\usepackage{amssymb}
\usepackage{amsmath}
\usepackage{xcolor}
\usepackage{hyperref}
\usepackage{amsmath}
\usepackage{amsfonts}

\hypersetup{colorlinks=true, linkcolor=blue, citecolor=blue, urlcolor=blue}

\long\def\comment#1{}

\begin{document}

\title{Search for post-inflationary QCD axions with a quantum-limited tunable microwave receiver}

\author{G. Sardo Infirri}\email[Corresponding author: ]{Giosue.Sardo@pd.infn.it} \affiliation{INFN, Sezione di Padova, Padova, Italy} \affiliation{Dipartimento di Fisica e Astronomia, Padova, Italy}
\author{D.~Alesini} \affiliation{INFN, Laboratori Nazionali di Frascati, Frascati, Roma, Italy}
\author{C.~Braggio} \affiliation{INFN, Sezione di Padova, Padova, Italy} \affiliation{Dipartimento di Fisica e Astronomia, Padova, Italy}
\author{G. Cappelli}\affiliation{Univ. Grenoble Alpes, CNRS, Grenoble INP, Institut N\'eel, 38000 Grenoble, France}
\author{G.~Carugno} \affiliation{INFN, Sezione di Padova, Padova, Italy} \affiliation{Dipartimento di Fisica e Astronomia, Padova, Italy}
\author{D.~D'Agostino} \affiliation{INFN, Sezione di Napoli, Napoli, Italy}
\author{A.~D'Elia} \affiliation{INFN, Laboratori Nazionali di Frascati, Frascati, Roma, Italy}
\author{D.~Di~Gioacchino} \affiliation{INFN, Laboratori Nazionali di Frascati, Frascati, Roma, Italy}
\author{R.~Di Vora}\affiliation{INFN, Laboratori Nazionali di Legnaro, Legnaro, Padova, Italy}
\author{M. Esposito} \altaffiliation[Present address: ]{CNR-SPIN Complesso di Monte S. Angelo, via Cintia, Napoli, 80126, Italy} \affiliation{Univ. Grenoble Alpes, CNRS, Grenoble INP, Institut N\'eel, 38000 Grenoble, France}
\author{P.~Falferi} \affiliation{Istituto di Fotonica e Nanotecnologie, CNR Fondazione Bruno Kessler, I-38123 Povo, Trento, Italy} \affiliation{INFN, TIFPA, Povo, Trento, Italy}
\author{U.~Gambardella} \affiliation{INFN, Sezione di Napoli, Napoli, Italy}
\author{A.~Gardikiotis} \affiliation{INFN, Sezione di Padova, Padova, Italy} 
\author{C.~Gatti} \affiliation{INFN, Laboratori Nazionali di Frascati, Frascati, Roma, Italy}
\author{C.~Ligi} \affiliation{INFN, Laboratori Nazionali di Frascati, Frascati, Roma, Italy}
\author{G. Lilli}\affiliation{INFN, Laboratori Nazionali di Legnaro, Legnaro, Padova, Italy}
\author{A.~Lombardi} \affiliation{INFN, Laboratori Nazionali di Legnaro, Legnaro, Padova, Italy}
\author{G.~Maccarrone} \affiliation{INFN, Laboratori Nazionali di Frascati, Frascati, Roma, Italy}
\author{D. Maiello}\affiliation{INFN, Sezione di Padova, Padova, Italy} \affiliation{Dipartimento di Fisica e Astronomia, Padova, Italy}
\author{A.~Ortolan} \affiliation{INFN, Laboratori Nazionali di Legnaro, Legnaro, Padova, Italy}
\author{A. Ranadive} \altaffiliation[Present address: ]{ Google Quantum AI, Goleta CA 93117, USA} \affiliation{Univ. Grenoble Alpes, CNRS, Grenoble INP, Institut N\'eel, 38000 Grenoble, France}
\author{A.~Rettaroli} \affiliation{INFN, Laboratori Nazionali di Frascati, Frascati, Roma, Italy} 
\author{N.~Roch}\affiliation{Univ. Grenoble Alpes, CNRS, Grenoble INP, Institut N\'eel, 38000 Grenoble, France}
\author{S.~Tocci} \affiliation{INFN, Laboratori Nazionali di Frascati, Frascati, Roma, Italy}
\author{G.~Ruoso}\email[Corresponding author: ]{Giuseppe.Ruoso@lnl.infn.it}\affiliation{INFN, Laboratori Nazionali di Legnaro, Legnaro, Padova, Italy}

\collaboration{QUAX collaboration}

\date{\today}


\begin{abstract}
A search for cosmological axions has been performed by scanning a frequency region of  $38\,$MHz centered at about $10.2\,$GHz, corresponding to an axion mass $m_a \simeq 42\,\mu$eV. The QUAX experimental apparatus, a haloscope comprised of a 1-liter volume tunable cavity immersed in an $8\,$T magnetic field and a quantum-limited detection chain,  set limits on the axion-photon coupling at the $10^{-14}\,$GeV$^{-1}$ level.
As no signal candidate has been observed, viable hadronic axion models are ruled out in a currently preferred post-inflationary  region $m_a > 40\,\mu$eV.
\end{abstract}

\maketitle


\textit{Introduction} -- 
The axion was introduced to solve the strong CP problem, i.e. the absence of CP violation in quantum chromodynamics (QCD).
Solution to the strong CP problem came with the new symmetry called Peccei Quinn (PQ), of which the axion is the pseudo Goldstone boson \cite{peccei1977cp,weinberg1978new,wilczek1978problem}.
The axion soon became a well-motivated dark matter candidate \cite{preskill1983cosmology,abbott1983cosmological,dine1983not}, but its mass is still unknown and the possible value ranges over several decades \cite{irastorza2018new}. 
Axion couplings to ordinary matter depend on the specific model used to incorporate the PQ symmetry into an extension of the standard model of particle physics.
The two most popular classes of models are known as KSVZ \cite{kim1979weak,shifman1980can} and DFSZ \cite{dine1981simple,zhitnitskij1980possible}.
A detailed description of their extended parameter space can be found in \cite{DILUZIO20201}.
With the assumption that all the local dark matter density is saturated by axions, a few experimental searches have been able to probe the specific axion-photon coupling predicted by such models.
The most sensitive probes are called haloscopes and are  based on tunable resonant cavities immersed in a strong magnetic field \cite{Sikivie:1983ip_halotheory,Sikivie:1985yu_halotheory}.
In the axion mass region below $30\,\mu$eV, only ADMX \cite{PhysRevD.109.012009} and CAPP \cite{PhysRevX.14.031023} experiments reached DFSZ sensitivity while  HAYSTAC \cite{PhysRevD.107.072007} achieved  near-KSVZ sensitivity. Searches at higher masses are flawed by unfavorable scaling of the cavity haloscope experimental parameters.
Recently, dish antenna \cite{PhysRevLett.134.171002}, dielectric \cite{garcia2024searchaxiondarkmatter} and plasma haloscopes \cite{PhysRevD.107.055013} were introduced to tackle higher masses, however the sensitivity of these instruments is still limited.

Recent works based on lattice simulations  \cite{Buschmann_2022,benabou2024axion}  motivate the search for post-inflationary axions \cite{PhysRevD.105.055025} with masses above 40 $\mu$eV.
In this mass region no extended searches have yet  been conducted at QCD axion sensitivity, required to probe the KSVZ or DFSZ classes of models.

The experiment QUaerere AXion (QUAX) has been conceived to search for axion dark matter  in the $35 - 45 \, \mu$eV mass window, corresponding to the frequency range from $8.5$ to $11\,$GHz. It is built around two complementary cavity haloscopes covering two distinct mass regions.
The QUAX haloscope  located in the Laboratori Nazionali di Frascati of INFN  is  designed to work in the lower half of the proposed mass range, and a pilot measurement has been recently performed at a mass of $\simeq 26.5 \, \mu$eV~\cite{PhysRevD.110.022008}.
The QUAX haloscope operating at the Laboratori Nazionali di Legnaro targets instead axion searches  above 10 GHz.
It has already conducted several test runs, obtaining near KSVZ model sensitivity in a region above 40$\,\mu$eV \cite{PhysRevD.99.101101,PhysRevD.103.102004,PhysRevD.106.052007,PhysRevD.108.062005}.

In this paper results are reported of a cosmological axion search using the improved high frequency QUAX haloscope.
The novel version of the apparatus allows for frequency tuning of the resonant detection system with uniform axion sensitivity over a broad range.
For the longest single acquisitions an axion sensitivity at about the KSVZ model is reached, while the typical sensitivity allows to probe axion-photon coupling values predicted within the hadronic QCD axion models \cite{DILUZIO20201}.


\textit{Experimental apparatus and data collection} -- 
The experimental setup is an evolution of the one presented in Ref. \cite{PhysRevD.108.062005}.
To allow for a wider frequency tuning, a new  dielectrically loaded cavity has been developed, in which a large effective volume is obtained by inserting a sapphire cylinder in a right cylindrical copper cavity and selecting the TM030 as the science mode.
The copper cylinder  is divided into two halves to allow for changing the frequency by a clamshell mechanism detailed in Ref. \cite{divora2024tunabledielectricresonatoraxion}, where a similar cavity working at about 11~GHz has been studied.
A computer controlled linear positioner acts on a pantograph that opens the two copper halves up to one degree, providing tuning from 10.212 GHz down to $10.154\,$GHz.
Larger tunings are  possible, but in this run the range was limited due to mechanical constraints. We expect to use the same cavity with increased tuning range in a forthcoming run. 

\begin{figure}[htb]
	
	\centering

	\includegraphics[width=0.45\textwidth]{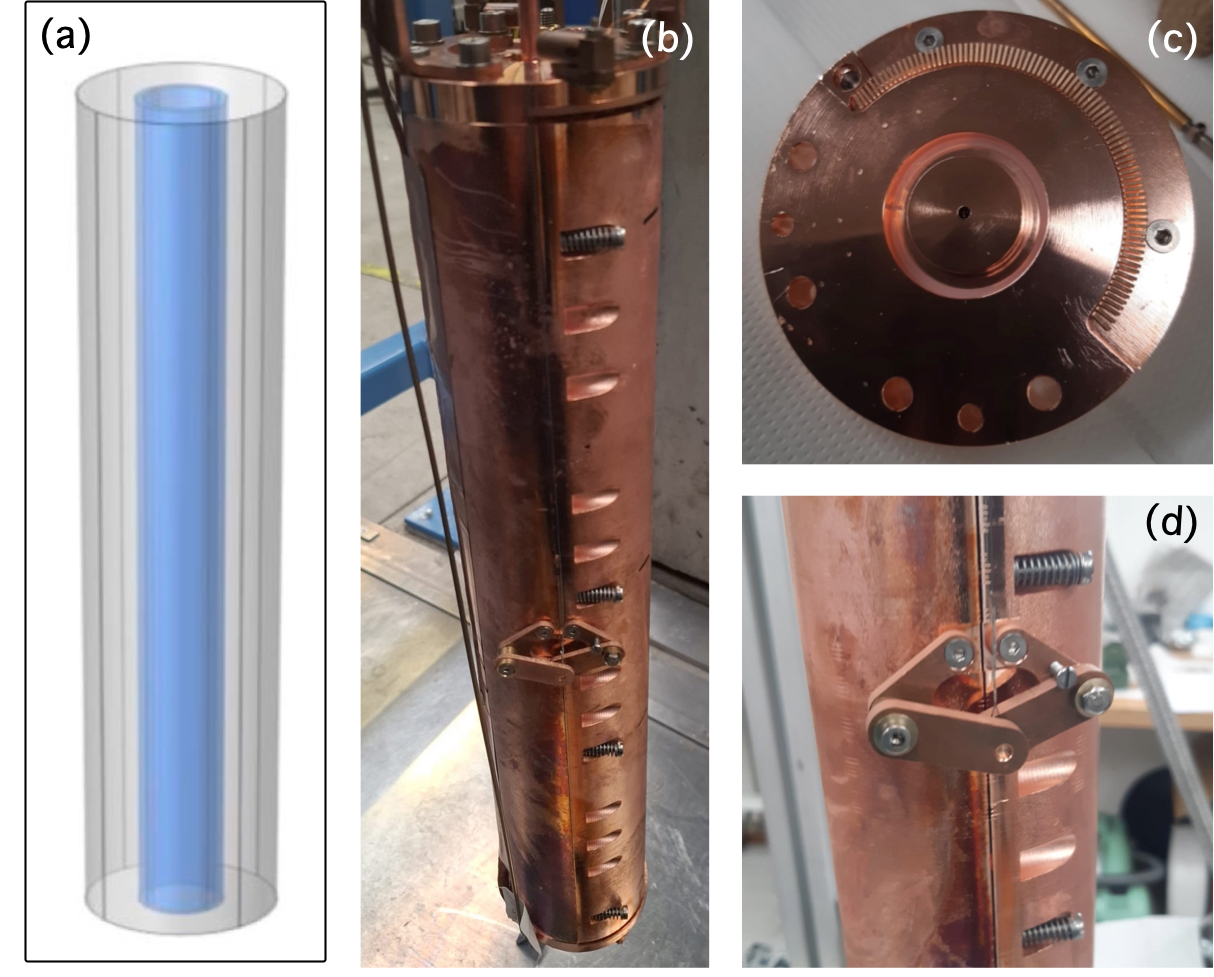}

	\caption{\small (a) The microwave resonator is a dielectrically loaded cylindrical cavity. The main structure is a copper cylinder with length $414.3\,$mm and $60.5\,$mm diameter (gray area). It includes a $420\,$mm long sapphire tube with $21.8\,$mm inner radius and $30.2\,$mm outer radius (blue area). The sapphire cylinder extends into the cavity endcaps by $5.3\,$mm deep annular grooves. (b) A picture of the cavity mounted in the dilution refrigerator is also shown. (c) End cap with annular grooves for holding the sapphire and springs for keeping the rf contact while tuning. (d) Pantograph for tuning.}

	\label{fig:Cavity}

\end{figure}

Detailed mode reconstruction is performed with finite element simulations to calculate the cavity form factor $C_{030}$ \cite{sikivie2021invisible}, which measures the coupling of the axion field to the specific cavity mode for different aperture.
Simulations give $C_{030}=0.43$, which complies with the room temperature bead pull measurement \cite{maier1952field}  giving $C_{030}=0.40\pm 0.02$.
This latter result is obtained with a cavity opening angle of about 0.3 degree.
Following results from simulations, this value is assumed to be constant throughout the frequency tuning region.

The apparatus lies inside the science chamber of a wet dilution refrigerator (DR), with the cavity kept at about $100\,$mK (see \cite{PhysRevD.108.062005} for details).
During data acquisition, the liquid helium level is monitored, and refills take place every 24 hours, thus ensuring good thermal stability of the system.
The cavity temperature is stable within a few mK, after a few days necessary for thermalization.
Temperature stability also ensures gain stability for the detection chain.

The detection chain is largely similar to what is described in  Ref.~\cite{PhysRevD.108.062005}, with a traveling wave parametric amplifier (TWPA)~\cite{Ranadive2022} used as a first stage amplifier. The TWPA, kept at about 120 mK, is followed by a cryogenic  high electron mobility transistor (HEMT) amplifier placed in the $4\,$K stage of the DR, and then by another HEMT at room temperature.

The output of the room temperature HEMT  is down-converted with a carrier frequency $f_{\rm LO}$ shifted by 1 MHz from the cavity frequency $f_c$.
In-phase and quadrature products are amplified, low pass filtered to avoid aliasing and sampled with an analog to digital converter working at the frequency of $4.4\,$MS/s.
Every $2^{23}$ samples, i.e. about $1.9\,$s,  data are stored in a file; in a typical acquisition step, 2000 data blocks of this length are acquired.

Two run sessions have been performed, the first in June 2024 and the second in November 2024.
The integrated acquisition time is about 225 hours, distributed over 18 days, with a resulting data collection efficiency of about $50\%$.
Residual time is mainly used for cavity tuning, optimization of the TWPA performance, and measurement of the working parameters: cavity resonance frequency and quality factor, antenna coupling, receiver gain profile and noise temperature at the cavity peak frequency.
Data acquisition is partially automated: while cavity tuning and amplifier optimization is still performed by the operator, all other measurements are controlled by the data acquisition software. 

\begin{figure}[htb]

  	\centering

	\includegraphics[width=0.46\textwidth]{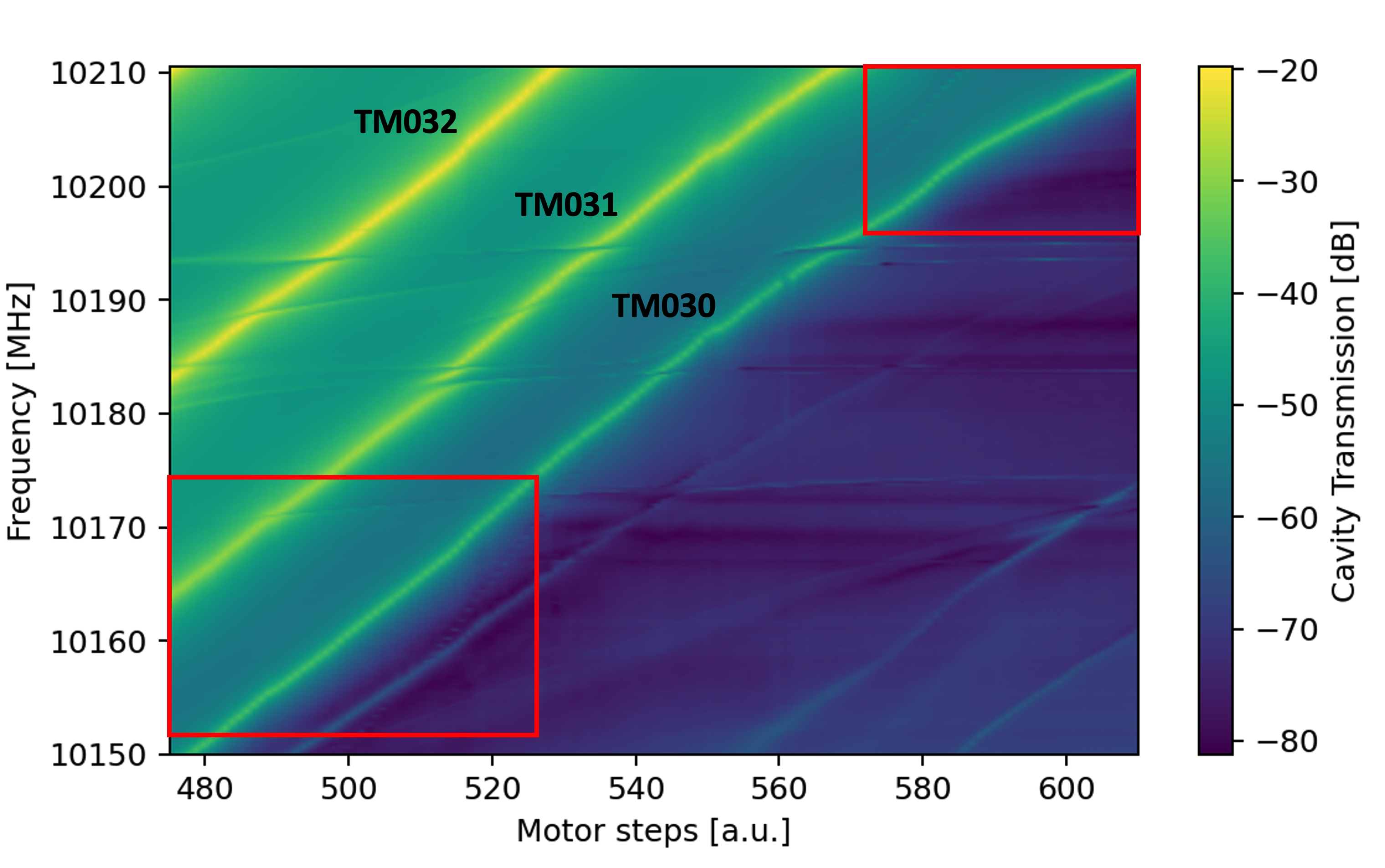}

	\caption{\small Cavity mode map showing the lowest order TM03n modes versus steps of the motor used for tuning. Science mode is the TM030, which can be tuned up to $10.212\,$GHz. Due to non-linearities in the actuation system, there is no strict linear relation between steps and frequency change. The studied frequency ranges are enclosed in the red boxes. It is also evident the presence of intruder modes in the region $10.180 - 10.196\,$GHz.  }

	\label{fig:Modemap}

\end{figure}

Axion searches  have been performed in a measurement window of about $38\,$MHz, corresponding to about 65\% of the available tuning range.
The presence of intruder modes (see Figure \ref{fig:Modemap}) degrading the system performance is the main reason for the reduced range, together with a mechanical problem in the tuning mechanism that precluded a fine tuning in the interval $10.175 - 10.196\,$GHz.
Tuning was performed in different step sizes, ranging from 1 to 2 cavity linewidths, for a total number of 124 measurement points $P_m$. 

\begin{figure}[htb]

	\centering

	\includegraphics[width=0.47\textwidth]{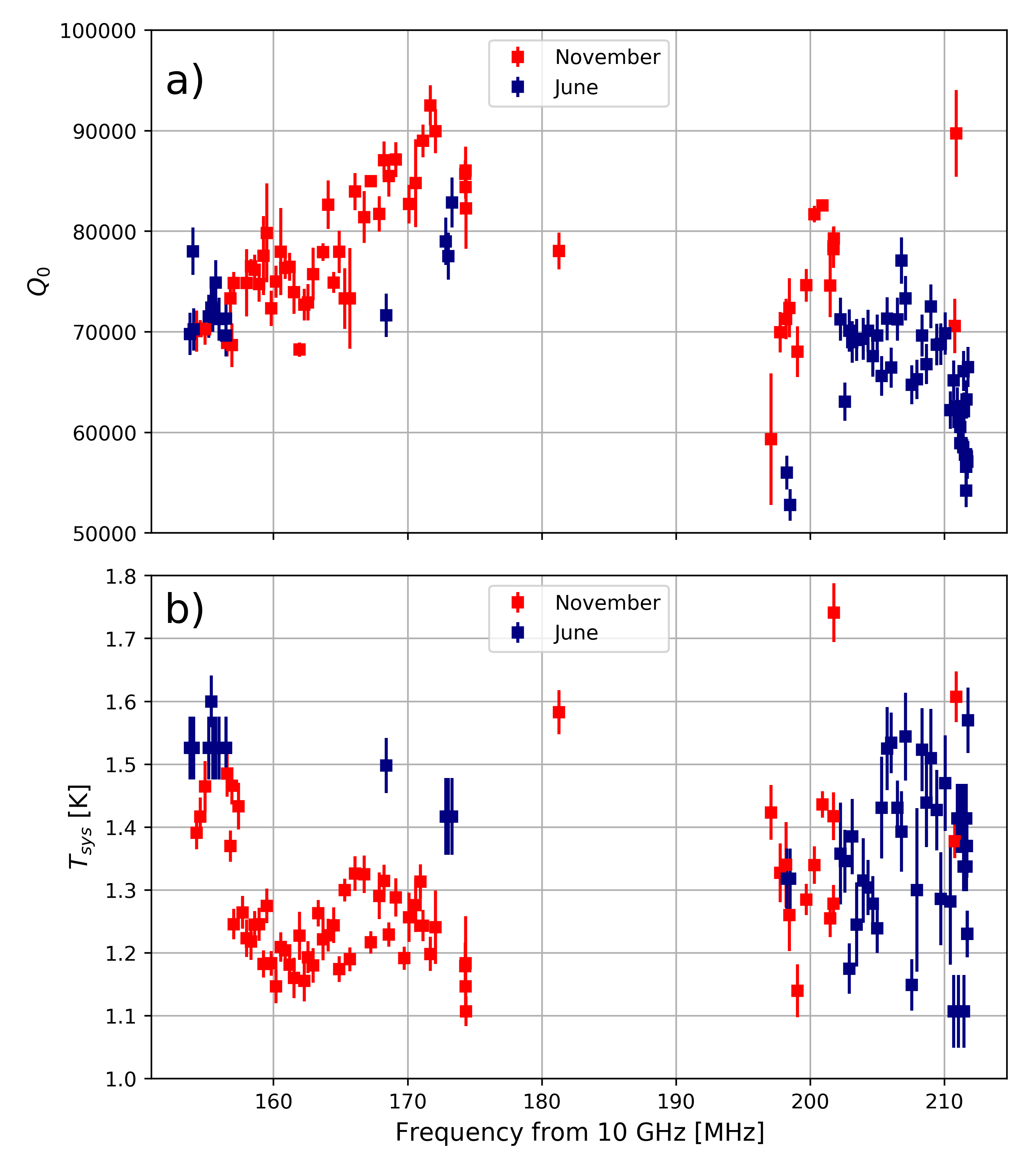}

	\caption{\small System parameters for each frequency measurement point $P_m$. Blue (red) data markers are from June (November) session. (a) Unloaded cavity quality factor $Q_0$. Cavity resonance frequency $f_c$,  coupling $\beta$, not shown in the figure, and quality factor $Q_0$ are derived from  transmission and reflection measurements using a spectrum analyzer with a tracking generator. (b) System noise temperature values $T_{sys}$.}

	\label{fig:Parameters}

\end{figure}

Figure \ref{fig:Parameters} (a) shows  the measured values of the cavity unloaded quality factor at each frequency tuning step.
Mechanical clearances cause different positioning of the cavity elements at each step, resulting in enhanced spreading of quality factor values.
Pick-up antenna coupling was set to a value just below the optimal value for scanning of $\beta =2$, and it was never optimized during the run.
Only a pair of data taking was completed with $\beta \simeq 1.1$ to check for better sensitivity. 

System noise temperature $T_{sys}$ is measured at each point $P_m$ using the procedure described in \cite{Braggio_2022}, improved compared to past versions thanks to a computer controlled system acting on a number of rf switches.
A single measurement of system noise temperature requires about $110\,$s.
The TWPA working parameters are optimized at every cavity frequency change: this optimization procedure is performed manually by setting the gain difference between having the pump driving the amplifier on or off above $25-26\,$dB.
This level was very close to the TWPA gain saturation: in the first session we used to keep a higher margin with respect to saturation, while in the second session we reduced this margin since we noticed that the TWPA was stable enough to allow for long-lasting measurements.
Since the added noise by the second stage amplifier is significant, higher TWPA gain resulted in lower values of $T_{\rm sys}$ \cite{Braggio_2022}.
The minimum observed $T_{\rm sys}$ was at about $1.1\,$K, corresponding to 2.3 photons (See Figure \ref{fig:Parameters} (b)).
This is the lowest noise temperature obtained for a tunable haloscope working above $40 \, \mu$eV \cite{cervantes2022admx,doi:10.1126/sciadv.abq3765,PhysRevLett.134.171002}.

The magnetic field is provided by a vertically aligned solenoidal coil of $450\,$mm length and $150\,$mm bore diameter.
With a current of $92\,$A the magnet delivers a peak magnetic field of $8.0\,$T.
The integrated value of $B^2$ over the cavity length is $50.2\,$T$^2$ m. A second coil reduces the stray field on sensitive electronics.
Moreover, the TWPA is shielded from residual magnetic fields by using two encapsulated boxes of lead and $\mu$-metal.

As mentioned above, typical measurements at points $P_m$ in axion search lasted about one hour.
Since the system was not completely automated, such steps were performed during daily operation of the haloscope.
During night operation, eight hours long data acquisitions   were performed.
Careful preliminary  analysis of the collected data showed that in most cases the working parameters did not change significantly during acquisition and that the sensitivity scaled as expected, i.e. following Dicke's formula \cite{dicke1946measurement} with the square root of integration time.
This is an important feature in case of positive alarms: longer integration time will increase the signal to noise ratio as envisaged.

\begin{figure*}[ht]

    \centering

    \includegraphics[width=0.99\textwidth]{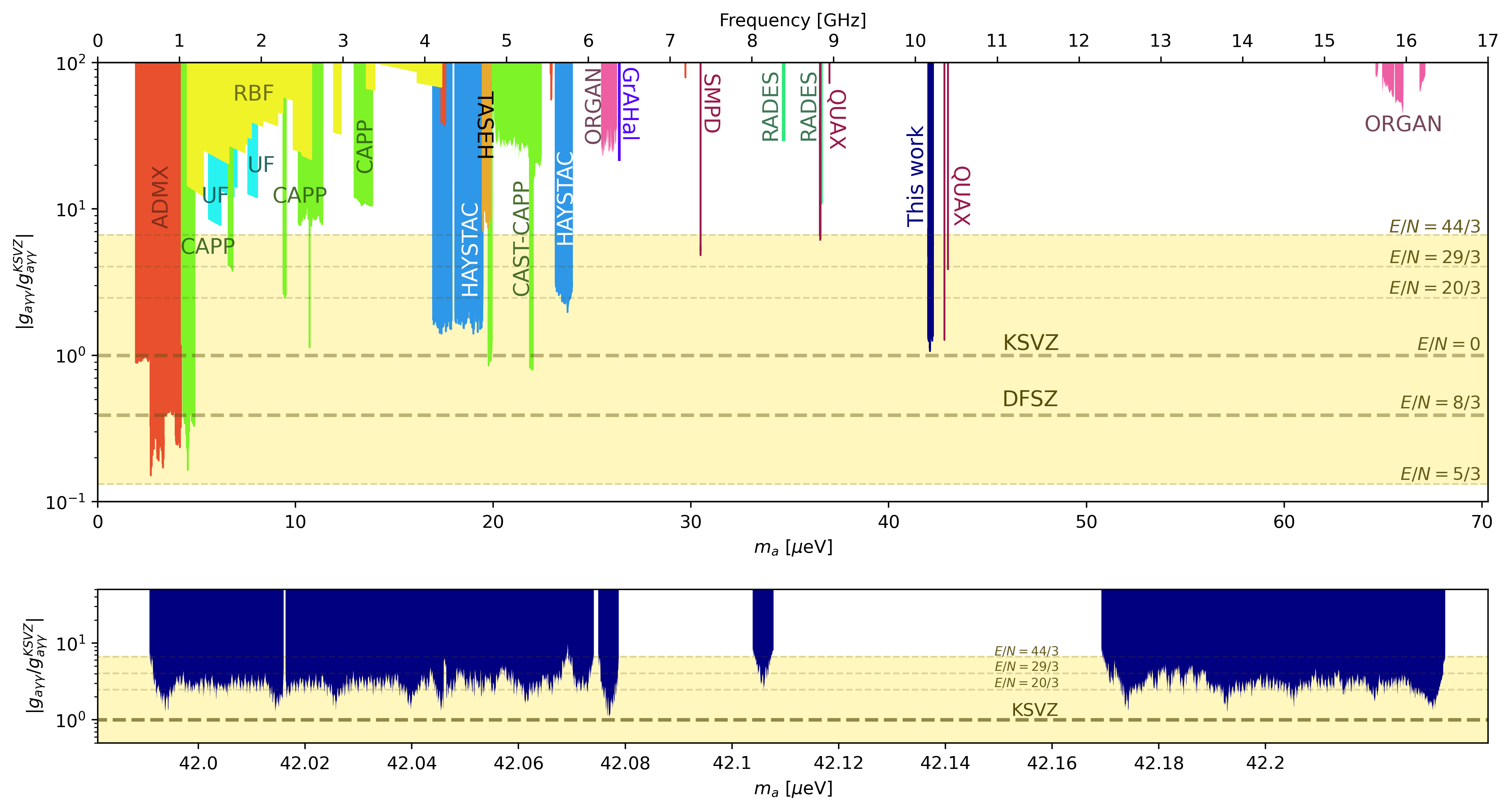}

    \caption{\small Axion–photon coupling exclusion regions at 90\% C.L.  as a function of axion mass from this work, previous QUAX results \cite{PhysRevD.99.101101,PhysRevD.103.102004,PhysRevD.106.052007,PhysRevD.108.062005,PhysRevD.110.022008} and other cavity haloscope searches:
    RBF~\cite{PhysRevLett.59.839,PhysRevD.40.3153},
    UF~\cite{PhysRevD.42.1297,HAGMANN1996209},
    ADMX~\cite{PhysRevLett.104.041301,PhysRevLett.120.151301,PhysRevLett.124.101303,PhysRevLett.127.261803,PhysRevLett.121.261302,PhysRevLett.134.111002,10.1063/5.0122907,PhysRevD.109.012009},
    CAPP~\cite{PhysRevLett.124.101802,PhysRevLett.125.221302,PhysRevLett.126.191802,PhysRevLett.128.241805,PhysRevD.106.092007,PhysRevLett.130.091602,PhysRevLett.130.071002,PhysRevLett.131.081801,PhysRevLett.133.051802,PhysRevX.14.031023},
    HAYSTAC~\cite{PhysRevLett.118.061302,Backes:2021aa,PhysRevD.97.092001,PhysRevD.107.072007,PhysRevLett.134.151006}, 
    TASEH~\cite{PhysRevLett.129.111802}, 
    CAST-CAPP~\cite{Adair:2022aa}, 
    RADES~\cite{Alvarez-Melcon:2021aa,ahyoune2025rades},
    ORGAN~\cite{doi:10.1126/sciadv.abq3765,PhysRevD.111.095007}, 
    and recent results using a single  microwave photon  detector (SMPD) \cite{PhysRevX.15.021031}.
    The yellow band encloses the expected axion photon coupling as predicted in phenomenologically preferred models for which the anomaly coefficient ratio $E/N \in (5/3,44/3)$ \cite{DiLuzio:2016sbl,DiLuzio:2017pfr}, generalisation of the standard KSVZ model ($E/N = 0  $)\cite{kim1979weak,shifman1980can}.
    Lower panel shows the closeup of the results of this work, in order to better identify the exclusion region for the models with $E/N = 44/3$ and $29/3$.
    The missing band at about $42.017\,\mu$eV is due to data removal for rf contamination.
    Plot generated using \cite{AxionLimits}.}

    \label{fig:results}

\end{figure*}


\textit{Data analysis and Results} -- 
Every saved file is divided in chunks of $2^{14}$ samples, corresponding to $\sim 4$ ms duration: on each of these chunks the complex Fast Fourier Transform is performed, with a resulting bin width of $\Delta \nu_{\rm bin} \simeq 268\,$Hz.
These down-converted power spectra (PSD) span the actual range [$f_{\rm LO} - 2.2 \,$MHz, $f_{\rm LO}+ 2.2 \,$MHz], with the cavity frequency belonging to  the upper half of each spectrum.

Some preprocessing of the data is due. Firstly, an unwanted intruding power at the mixer level, which slightly affects most of the PSDs in a small fraction of the total bins, is removed.
This was caused by a defective power supply of the HEMT amplifier.
The disturbance is symmetric about $f_{LO}$ and is removed from the upper half of each PSD using lower half values.

The stability of the detection chain gain and of the cavity resonance frequency at each measurement $P_m$ are checked. The former is estimated by injecting a low power pure tone  in the detection chain at a frequency $f_{\rm ref}=f_{LO}- 200\,$kHz.
The measured power at $f_{\rm ref}$, obtained for all blocks of 512 consecutive PSDs (about $1.9\,$s) of a specific $P_m$, is used to normalize all its PSDs.
Another required normalization takes into account frequency-dependent contributions of the amplification chain, namely the TWPA gain profile and the transmissivity of the ADC input filters.

The stability of the cavity resonance frequency can be checked by fitting the power spectra in a region around $f_c$ with a Fano like function \cite{PhysRevD.108.062005,PhysRevD.110.022008}.
This is performed on the average of $20\times 512$ consecutive PSDs, corresponding to a $38\,$s time window.
In case of variations of the resonance frequency above 20\% of the cavity linewidth in a point $P_m$, varied data are moved to a new measurement point.
This process increases the number of measured points to 147, each having a corresponding averaged spectrum, with proven constant cavity resonance frequency, quality factor and antenna coupling.  

For each averaged spectrum a frequency window of maximum three cavity linewidths inside the interval $\left[f_{\rm LO} + 0.5\,{\rm MHz}, f_{\rm LO} + 1.5\,{\rm MHz}\right]$ is selected.
To recover the appropriate power units, the spectra are normalized, at the cavity peak $f_c$, to the power given by the noise temperature, i.e. $k_B T_{\rm sys} \Delta \nu_{\rm bin}$, where $k_B$ is the Boltzmann constant.
The spectra are then divided  by the cavity shape to obtain an input noise spectrum $S_m (f)$ for each $P_m$, where it is possible to look for an added axion signal.

For each $S_m (f)$, a Savitzky-Golay (SG) filter, with polynomial order 4 and window range $\sim 110\,$kHz,  is used to estimate the baseline $S_{\text{bl,m}} (f)$.
To search for axion excess power, confidence belts (CBs) are computed using Monte Carlo simulations. Axion signal power is given by:
\begin{equation}
    P_{\rm in}^a(\nu, \nu_a)= g_{a\gamma\gamma}^2 \frac{\hbar^3 c^3\rho_a}{m_a^2}\frac{2 \pi \nu_a}{\mu_0} f_a(\nu, \nu_a)
    \label{eq:axion_power}
\end{equation}
where $g_{a\gamma\gamma}$ is the coupling constant of the axion-photon interaction, $m_{a} = h \nu_a$ is the axion mass, $\rho_a \sim 0.45$\,GeV/cm$^3$ is the local dark matter density~\cite{10.1093/ptep/ptaa104} with Maxwell-Boltzmann distribution  $f_a(\nu, \nu_a)$, characterised by a width $\delta_{\nu_a} \sim 10\,$kHz in the Earth reference frame~\cite{turner1990periodic}.

The entire measurement window is analyzed at the hypothetical frequencies $\nu_a^i$, spaced by $\delta_{\nu_a}/2$.
Assuming the Dicke radiometer equation and using as baseline $S_{\text{bl,m}} (f) + k \cdot P_{\rm in}^a(\nu, \nu_a^i)$, $10\,000$ random spectra have been generated for each $\nu_a^i$ and all $k$.
The values of $k$ are chosen such that the signal to noise ratio (SNR) ranges from $0$ to $10$ at steps of $0.25$.
SNR = 1 occurs when  $k \cdot P_{\rm in}^a(\nu, \nu_a^i)$  integrated in $I_a = [\nu_a^i, \nu_a^i + \delta_{\nu_a}]$ is equal to the variance of the noise in the same interval.
The chosen interval is proven to maximize the SNR for the axion lineshape in Eq. (\ref{eq:axion_power}).
Axion power is calculated using $g_{a\gamma\gamma}^{\rm KSVZ} (\nu_a \simeq 10.2\, {\rm GHz})\simeq 1.67 \times 10^{-14}\,$GeV$^{-1}$, the coupling constant of KSVZ model \cite{kim1979weak,shifman1980can}.
These data are used for the construction of the CBs after a baseline removal.
Every CB is corrected to take into account the excess power non-negativity \cite{feldman1998unified}.
All the CBs show that the  SG filtering is responsible for a $\sim 30 \%$ underestimation of the axion power \cite{rosso2025baseline}.

In order to have an expected false alarm rate of one per month, the detection threshold is set to $\text{SNR} = 4.5$ using receiver operating characteristic curves.
Since no rescan candidate is found, the PSDs are used to set a $90 \%\,$C.L. limit on the coupling constant $g_{a\gamma \gamma}$ using Eq.~(\ref{eq:axion_power}).
Figure \ref{fig:results} shows the resulting limits, which include the error budget for the measured cavity and receiver parameters, accounting on average about $4 \%$.
Hadronic axion models with anomaly coefficient ratio $E/N = 44/3$ ($E/N = 29/3$) are excluded in a mass interval of $153$ neV ($136$ neV), located between $41.991\,\mu$eV and $42.234\,\mu$eV \cite{DILUZIO20201}.
As a global characterization of the apparatus, a 90\%  C.L. limit sensitivity of about $\left|g_{a\gamma\gamma}\right| \simeq 3.4 \times \left|g_{a\gamma\gamma}^{\text{KSVZ}}\right|$ over a $300\,$kHz band is reached on average with an integration time of one hour.


\textit{Conclusion and Outlook} -- 
The scanning within a $\sim 60\,$MHz extended region with a sensitivity to axion-photon coupling up to the KSVZ class of models presented in this Letter is a significant advancement in the search for dark matter axions in the well motivated region  $m_a \gtrsim 40 \,\mu$eV. 
This outcome relies on the use of a dielectrically loaded tunable cavity designed to have large effective volume even at high frequency, and on a quantum-limited, wide-bandwidth detection chain that we successfully operated in the presence of a strong magnetic field.
Moving forward, the QUAX apparatus will be enhanced by a more powerful magnet, and in addition a duty cycle increase is foreseen, thus heading for  better sensitivity and  faster scan rate.
This apparatus lays the foundations for a tunable haloscope capable to cover the proposed frequency range from $9.5$ to $11\,$GHz.


\hfill \break

We are grateful to E. Berto, A. Benato, and M. Rebeschini for the mechanical work; F. Calaon and M. Tessaro for help with the electronics and cryogenics.
We thank G. Galet and L. Castellani for the development of the magnet power supply, and M. Zago who realized the technical drawings of the system.
We thank J. Pazzini, F. Fanzago and A. Calanca for setting up the cloud data storage.
We thank L. Di Luzio for support on hadronic axion models.
We deeply acknowledge the Cryogenic Service of the Laboratori Nazionali di Legnaro for providing us with large quantities of liquid helium on demand. 

This work is supported by INFN (QUAX experiment), by the U.S. Department of Energy, Office of Science, National Quantum Information Science Research Centers, Superconducting Quantum Materials and Systems Center (SQMS) under Contract No. DE-AC02-07CH11359.
Optimization of the TWPA amplifier was accomplished within the INFN project PNRR-ICSC Spoke 10.

\appendix

\bibliography{PRLRun.bib}

\end{document}